\documentclass{article}

\usepackage{arxiv}

\usepackage[utf8]{inputenc} % allow utf-8 input
\usepackage[T1]{fontenc}    % use 8-bit T1 fonts
\usepackage{hyperref}       % hyperlinks
\usepackage{url}            % simple URL typesetting
\usepackage{booktabs}       % professional-quality tables
\usepackage{amsfonts}       % blackboard math symbols
\usepackage{nicefrac}       % compact symbols for 1/2, etc.
\usepackage{microtype}      % microtypography
\usepackage{lipsum}		% Can be removed after putting your text content
\usepackage{graphicx}
\usepackage{natbib}
\usepackage{doi}
\usepackage{epic,eepic}
\usepackage{bm,enumerate,amsmath,amssymb}
\usepackage{epsfig}
\usepackage{float}
\usepackage{subfigure}

\title{Operational modal analysis of under-determined system based on Bayesian CP decomposition}

\author{ \href{https://orcid.org/0000-0001-7607-3028}{\includegraphics[scale=0.06]{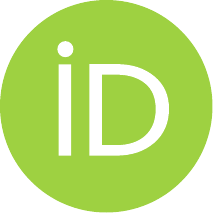}\hspace{1mm}Sunao TOMITA} \\
	Toyota Central R \& D Labs., Inc.\\
	Bunkyo-ku, Tokyo,  112-0004, Japan \\
	\texttt{stomita@mosk.tytlabs.co.jp} \\
	\And
        \href{https://orcid.org/0000-0001-6863-4868}
 	{\includegraphics[scale=0.06]{orcid.pdf}\hspace{1mm}Tomohiko JIMBO} \\
	Toyota Central R\&D Labs., Inc.\\
	Yokomichi, Nagakute, Aichi, 480-1192, Japan \\
	\texttt{t-jmb@mosk.tytlabs.co.jp} \\
}

\begin{document}
\maketitle

\begin{abstract}
Modal parameters such as natural frequencies, modal shapes, and the damping ratio are useful to understand structural dynamics of mechanical systems. Modal parameters need to be estimated under operational conditions for use in structural health monitoring. Therefore, operational modal analysis (OMA) without input signals has been proposed to easily extract modal parameters under operational conditions. Recently, OMA for under-determined systems with more active modes than measurement outputs has been investigated to reduce the number of sensors. This study proposes the OMA framework for under-determined systems based on Bayesian CP (CANDECOMP/PARAFAC) decomposition of second-order statistical data. The proposed method enables us to extract the modal parameters from under-determined systems without tuning the number of active modes, because the rank of the tensor data corresponding to the number of active modes is automatically determined via Bayesian inference. The effectiveness of this method is demonstrated using artificial vibration data of a mass-spring system under operational and under-determined conditions.
\end{abstract}

% keywords can be removed
\keywords{Operational modal analysis, Modal properties extraction, Blind source separation, Tensor decomposition, CANDECOMP/PARAFAC decomposition, Bayesian inference, Under-determined problem, Second-order blind identification}

\section{\label{intro}Introduction}
Considering structural aging, health monitoring technology is crucial to ensure safety of the society. However, health monitoring is currently costly and manpower-intensive, and its automation is an important issue for the aging society. In health monitoring, modal parameters such as natural frequencies of mechanical structures are crucial in detecting abnormal signals caused by failures. Generally, experimental modal analysis is widely used where the input is known and the transfer characteristics between the input and output are used to estimate the modal parameters (Nagamatsu, 1993). Contrastingly, as experimental modal analysis is expensive in obtaining inputs, operational modal analysis (OMA), which estimates modal information using only the output signals of a mechanical system under operational conditions, has been proposed (James et al., 1995).

Recently, blind source separation (BSS) has been applied for OMA. For instance, independent component analysis (Hyvärinen and Oja, 1997) is used in OMA (Kerschen et al., 2007), Second-Order Blind Identification is also applied in OMA  (Poncelet et al., 2007; Zhou and Chelidze, 2007; McNeill and Zimmerman, 2008). Using BSS in OMA is advantageous in that it reduces the number of tuning parameters. However, the application of classical BSS is limited to determined or over-determined systems (i.e., number of sensors is the same or larger than than number of vibration modes). Therefore, it is important to develop OMA for under-determined systems where the number of sensors is smaller than the number of active vibration modes in the system.

The tensor decomposition approach called CP (CANDECOMP/PARAFAC) or PARAFAC (parallel factors) decomposition (Kolda and Bader, 2009)  enables us to realize BSS for under-determined systems (De Lathauwer and Castaing, 2008). BSS using CP decomposition decomposes the three-dimensional data (i.e., tensor data) comprising covariance matrices for multiple time-delayed signals into a sum of rank-1 tensors to identify more input signals (i.e., source signals) than the number of output signals (i.e., sensors). This technique has been applied to OMA with a limited number of sensors (Abazarsa et al., 2013; Friesen and Sadhu, 2017). Similar to tensor decomposition, sparse estimation has also been applied for BSS of under-determined systems, which has been used for OMA (Yang and Nagarajaiah, 2013; Yu et al., 2014). Furthermore, a method for solving under-determined OMA for a distributed set of sensors has been proposed by combining the sparse estimation framework and tensor decomposition (Sadhu et al., 2013). Several previous studies have proposed OMA for under-determined systems using tensor decomposition. Moreover, the CP decomposition in these studies requires determining the CP ranks corresponding to the number of active modes through trial and error before identification, which makes it difficult to automate OMA. Although methods using singular values (Abazarsa et al., 2013) or selecting an appropriate CP rank from multiple-rank CP decomposition results (Sadhu et al., 2015) have been proposed, determination of CP rank has not been fully established.

CP rank determination has also been studied in the machine learning community (Mørup and Hansen, 2009). In particular, Zhao et al. recently proposed a method for determining CP rank using Bayesian CP factorization (BCPF) (Zhao et al., 2015). This method is superior to conventional CP decomposition considering the following four points: (1) The CP rank can be determined automatically. (2) No tuning parameters are required. (3) Probabilistic information is available. (4) It converges empirically quickly. These features of BCPF enable the automation of OMA of under-determined systems. However, to the best of our knowledge, previous studies have not applied BCPF to OMA of under-determined systems.
	
In this study, we propose an OMA using BCPF of three-dimensional tensors comprising covariance matrices with multiple time delays in order to automatically estimate the number of modes of OMA for under-determined systems. The BCPF facilitates the automation of OMA, since the number of vibration modes is estimated as CP ranks in the process of CP decomposition of the three-dimensional tensors.
	
This paper is organized as follows. First, we briefly review how BSS can be applied to OMA based on the similarity between BSS and OMA, and explain the BSS based on CP decomposition that can be applied for tensor decomposition of under-determined systems. Next, we present the formulation of OMA using BCPF and propose an OMA for under-determined systems that does not require tuning parameters. In addition, we apply the proposed method to numerical experimental data of the under-determined system and verify that the number of modes is automatically determined and modal parameters are properly estimated. Finally, we present conclusions.

	\section{\label{conventional} Operational modal analysis of under-determined systems by blind source separation}
	This section briefly explains OMA for under-determined systems based on the CP decomposition. First, the correspondence between the BSS and  OMA is shown, and the applicability of the BSS algorithm to OMA is demonstrated. Next, BSS based on second-order statistics is explained as the application of BSS to OMA. Finally, the theory of estimating more vibration modes than the number of sensors by CP decomposition of the second-order statistics is explained.
	
	\subsection{Operational modal analysis and blind source separation}
	The similarities between the problem setting of BSS and OMA are explained. First, a theoretical modal analysis of a structure is described as the problem setup of OMA. The equations of motion for a structure with $N$ degrees of freedom are:
	\begin{equation}
		\mathbf{M\ddot{x}}+\mathbf{C\dot{x}}+\mathbf{Kx}=\mathbf{f},
		\label{eq:motion}
	\end{equation}
	\noindent
	where, $\mathbf{M}\in {{\mathbb{R}}^{N\times N}}$, $\mathbf{K}\in {{\mathbb{R}}^{N\times N}}$ and $\mathbf{C}\in {{\mathbb{R}}^{N\times N}}$ are the mass, the stiffness, and damping matrices, respectively. $\mathbf{x}\in {{\mathbb{R}}^{N}}$ is displacement vector and $\mathbf{f}\in {{\mathbb{R}}^{N}}$ is force vector. Assuming proportional damping, modal decomposition of Eq. (\ref{eq:motion}) leads to:
	\begin{equation}
		\mathbf{x}\left( t \right)={\mathbf{\Phi }}\bm{\eta}(t),
		\label{eq:ModalDecomposition}
	\end{equation}
\noindent
where, $\mathbf{\Phi }={{[\bm{\phi} _{1}^{{}},...,\bm{\phi} _{i}^{{}},...,\bm{\phi} _{N}^{{}}]}^{\mathrm{T}}}$ are normal modes of the proportional damping systems. Each column of the modal vector $\bm{\phi}  _{i}$ is vibration modes. The modal responses $\bm{\eta }={{[\eta _{1}^{{}},...,\eta _{i}^{{}},...,\eta _{N}^{{}}]}^{\mathrm{T}}}$ are:
\begin{equation}
	{{\eta }_{i}}={{u}_{i}}{{e}^{-{{\zeta }_{i}}{{\omega }_{i}}t}}\cos \left( {{\omega }_{i}}t+{{\theta }_{i}} \right),
	\label{eq:ModalResponse}
\end{equation}
\noindent 
where, $\omega_i$ and $\zeta_i$ are the natural frequencies and damping ratios, respectively. $u_i$ and $\theta_i$ are coefficients determined by the initial states. The OMA identifies the natural frequencies $\omega$, damping ratios $\zeta$, modal responses $\eta$, and modal vectors $\bm{\phi}$ for $N$-vibration modes using only $M$-output signals $\mathbf{x}$.\par

In contrast, BSS assumes mixture models defined by $M$-output signals $\mathbf{x}(t)\in {{\mathbb{R}}^{M}}$ and $N$-input signals (source signals) $\mathbf{s}(t)\in {{\mathbb{R}}^{N}}$:
\begin{equation}
	\mathbf{x}(t)=\mathbf{{A}}\mathbf{s}(t).
	\label{eq:BSS}
\end{equation}

\begin{figure}[h]
%			\begin{tabular}
	\begin{center}

		\subfigure[Blind source separation]{
			\includegraphics[clip, width=0.5\linewidth]{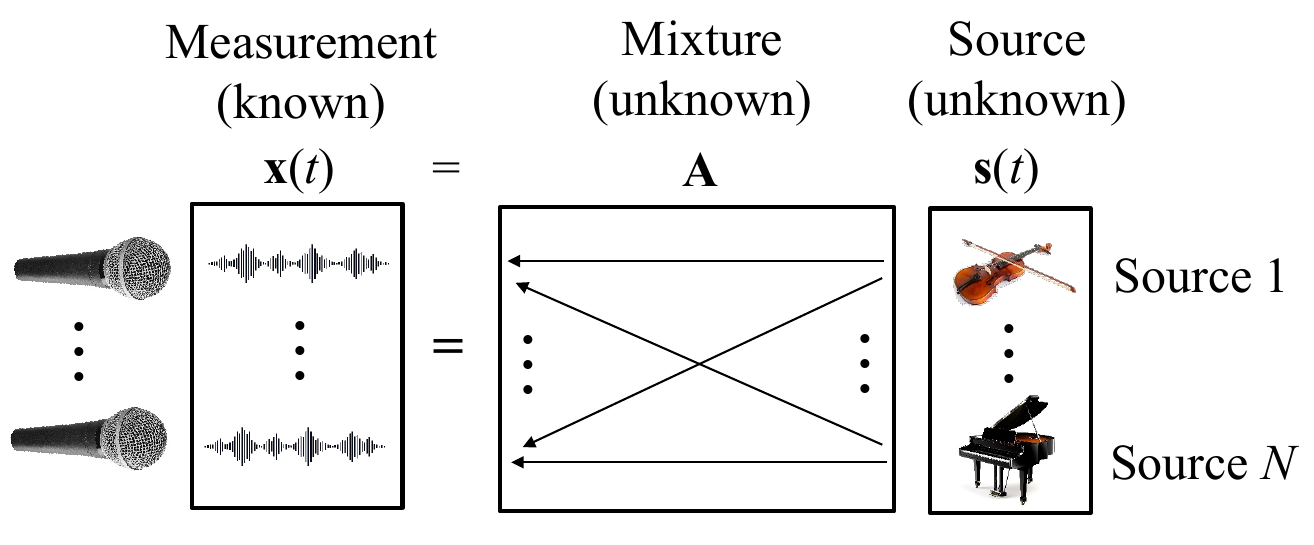}
			\label{fig:BSS}
		}
		\subfigure[Operational modal analysis]{
			\includegraphics[clip, width=0.5\linewidth]{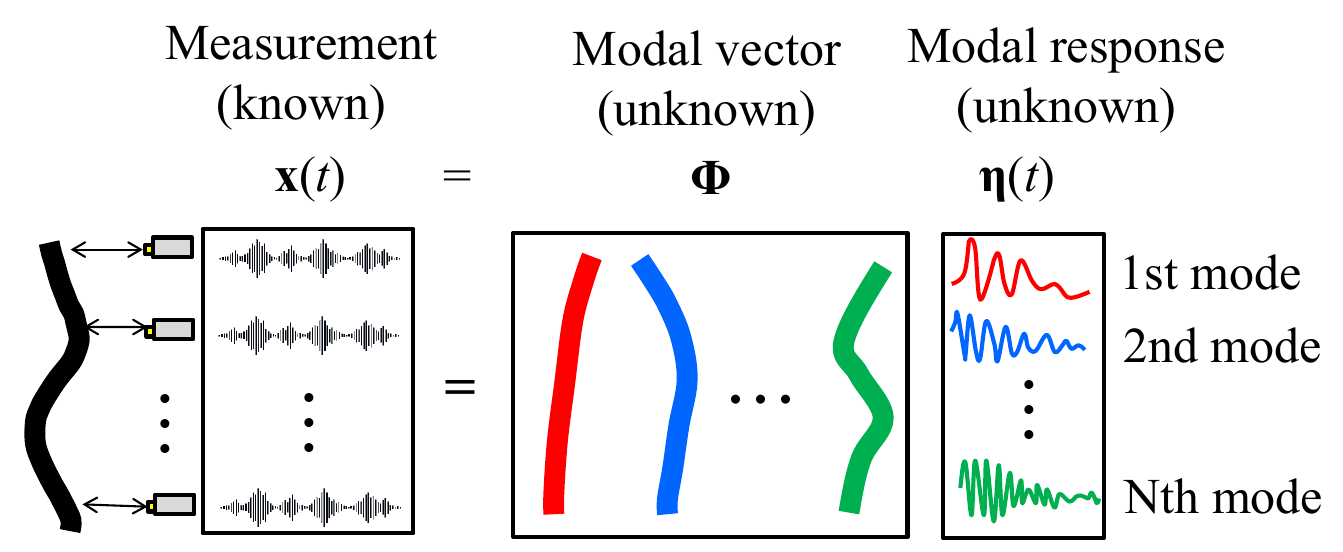}
			\label{fig:OMA}
		}
	\end{center}
%		\end{tabular}
	%	\vspace{-0.5cm}
	\caption{Similarity between blind source separation and operational modal analysis. This similarity enables us to apply blind source separation for operation modal analysis.}
	\label{fig:BSS_OMA}
\end{figure}

\noindent
Only $M$-output signals $\mathbf{x}$ are used to identify the mixture matrix $\mathbf{{A}}\in {{\mathbb{R}}^{M\times N}}$ and $N$-input signals $\mathbf{s}$. The mixture matrix $\mathbf{{A}}$ and input signals $\mathbf{s}$ corresponds to modal vectors $\mathbf{\Phi}$ and modal responses $\bm{\eta}$ in theoretical modal analysis (Eq. (\ref{eq:ModalDecomposition}), as illustrated in Fig. \ref{fig:BSS_OMA}.

%%BSSの説明

Therefore, the BSS frameworks can be applied for OMA to identify the vibration vectors $\mathbf{\Phi }={{[\bm{\phi} _{1}^{{}},...,\bm{\phi} _{i}^{{}},...,\bm{\phi} _{N}^{{}}]}^{\mathrm{T}}}$ and modal responses $\bm{\eta }={{[\eta _{1}^{{}},...,\eta _{i}^{{}},...,\eta _{N}^{{}}]}^{\mathrm{T}}}$. Furthermore, the eigenvalues are identified by the Fourier transforms of the modal responses:
\begin{equation}
	{{s}_{i}}=-{{\zeta }_{i}}{{\omega }_{i}}{\pm}{{\zeta }_{i}}{\mathrm{j}}{\omega }_{i}\sqrt{1-{\zeta_i }_{1}},
	\label{eq:pole}
\end{equation}
where $\mathrm{j}$ is the imaginary unit. The natural frequencies and damping ratios are obtained from the following equation:
\begin{equation}
	{f _{i}}=\frac{|s_i|}{2\pi},\ 
	{{\zeta}_{i}}=-\frac{\mathrm{Re}(s_i)}{|s_i|}.
	\label{eq:poletomodal}
\end{equation}
\noindent

%%%%%%%%%%%%%%%%%%%%%%%%%%%%%%
\subsection{Blind source separation using second-order statics}
This study employs the second-order blinds source separations as the BSS algorithm, which is briefly reviewed in this subsection. Assuming that input signals $\mathbf{s}(t)$ and delayed signals $\mathbf{s}(t+\tau)$ shifted by time $\tau$ are uncorrelated, the covariance matrix of these signals ${\tilde{\mathbf{{R}}}_{\mathbf{s}}}\left( \tau  \right)\in {{\mathbb{R}}^{N\times N}}$ is:
%%%
\begin{equation}
	{\tilde{\mathbf{{R}}}_\mathbf{s}}\left( \tau  \right)=\mathbb{E}\left[ \mathbf{s}\left( t+\tau  \right)\mathbf{s}{{\left( t \right)}^{\mathrm{T}}} \right]=\mathrm{diag}\left[ {{\rho }_{1}}\left( \tau  \right),...,{{\rho }_{N}}\left( \tau  \right) \right],\
	\label{eq:convariance matrix of input signals}
\end{equation}

\noindent
where, $\rho$ is the auto-covariance of the input signals. Therefore, eigenvalue problems can be obtained from the covariance matrix ${\tilde{\mathbf{{R}}}_{\mathbf{x}}}\left( {{\tau }_{k}} \right)\in {{\mathbb{R}}^{M\times M}}$ as:
\begin{equation}
	{\tilde{\mathbf{{R}}}_{\mathbf{x}}}\left( {{\tau }_{k}} \right)=\mathbb{E}\left[ \mathbf{x}\left( t+{{\tau }_{k}} \right)\mathbf{x}{{\left( t \right)}^{\mathrm{T}}} \right]=\mathbf{{A}}{\tilde{\mathbf{{R}}}_{\mathbf{s}}}\left( {{\tau }_{k}} \right){{\mathbf{{A}}}^{\mathrm{T}}},
	\label{eq:convariance matrix of output signals}
\end{equation}
%\color{red}
where, ${\tau}_{k}(k=1,...,K)$ is the $k$-th time delay, T is the transpose of a matrix, and  $\mathbb{E}$ is the expectation value. In BSS based on JADE (joint approximate diagonalization of eigenmatrice, Belouchrani et al., 1997) that simultaneously diagonalizes all covariance matrices. Output signals $\mathbf{z}$ are whitened as:
\begin{equation}
	\mathbf{z}
	=\mathbf{W}\mathbf{x}
	={\tilde{\mathbf{{R}}}_{\mathbf{s}}}^{-1/2}\mathbf{{A}}\mathbf{x},
	\label{eq:whitening}
\end{equation}
%\color{black}
by multiplying the regular matrix $\mathbf{W}$ to output signals $\mathbf{x}$. Then, JADE can be applied for only determined ($M=N$) or over-determined systems ($M>N$).

%%%%%%%%%%%%%%%%%%%%%%%%%%%%%%%%
\subsection{Blind source separation based on CP decomposition}
\begin{figure}[H]
	\begin{center}
		\includegraphics[clip, width=1\linewidth]{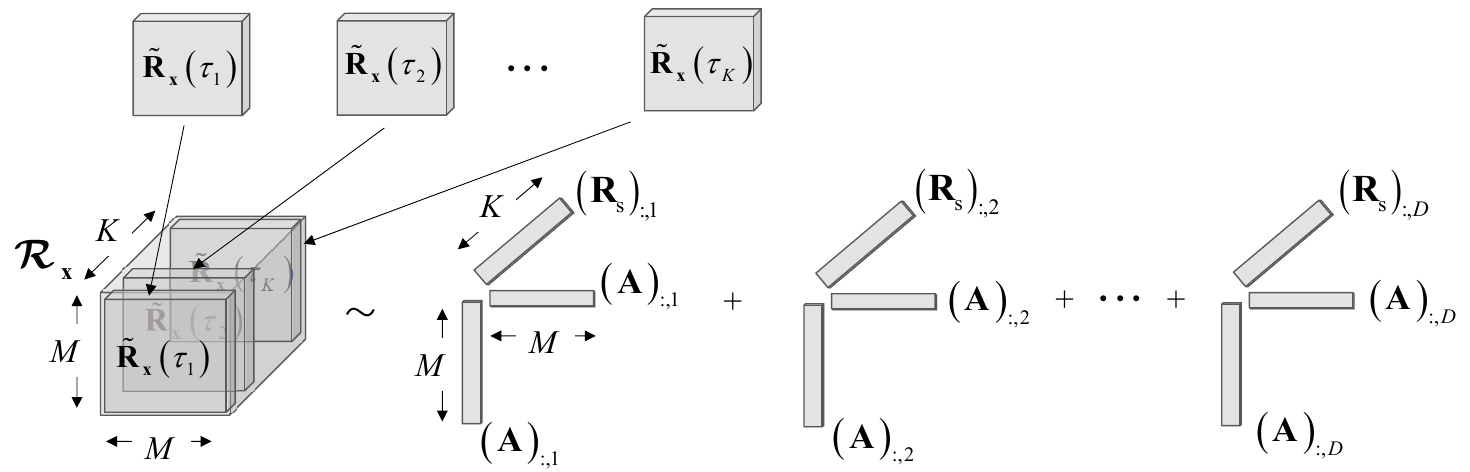}
	\end{center}
	\caption{Schematics of simultaneous diagonalization through CP decomposition of three-dimensional tensors of covariance matrices.}
	\label{fig:CPD}
\end{figure}

To simultaneously diagonalize the covariance matrices for under-determined systems, CP decomposition is employed (De Lathauwer and Castaing, 2008), where the uniqueness of the CP decomposition enables us to apply BSS to the under-determined systems. First, using the covariance matrix ${\tilde{\mathbf{{R}}}_{\mathbf{x}}}\left( {{\tau }_{k}} \right)\in {{\mathbb{R}}^{M\times M}}$ of output signals $\mathbf{x}(t)$ with time delays $\tau_k(k=1,2,...,K)$ shown in Eq. (\ref{eq:convariance matrix of output signals}), the three-dimensional tensor ${{{\bm{\mathcal{R}}}_{\mathbf{x}}}}\in {{\mathbb{C}}^{M\times M\times K}}$:
\begin{equation}
	{{\left( {{{\bm{\mathcal{R}}}_{\mathbf{x}}}} \right)}_{ijk}}:={{\left( {\tilde{{\mathbf{{R}}}}_{\mathbf{x}}}({{\tau }_{k}}) \right)}_{ij}}
	\label{eq:definition of tensor},
\end{equation}
is defined and decomposed as a summation of $D$-rank-one tensors as:
\begin{equation}
	{{\left( {{{\bm{\mathcal{R}}}_{\mathbf{x}}}} \right)}_{ijk}}=\sum\limits_{r=1}^{D}{{{\left( \mathbf{A} \right)}_{ir}}{{\left( \mathbf{A} \right)}_{jr}}{{\left( {{\mathbf{R}}_{\mathrm{s}}} \right)}_{kr}}}
	\label{eq:CPdecomposition_1}.
\end{equation}
The schematic of CP decomposition of the covariance matrix is shown in Fig. \ref{fig:CPD}  and defined as
\begin{equation}
	{{\bm{\mathcal{R}}}_{\mathbf{x}}}=\sum\limits_{r=1}^{D}{{{\left( \mathbf{A} \right)}_{:,r}}\circ {{\left( \mathbf{A} \right)}_{:,r}}\circ {{\left( {{\mathbf{R}}_{\mathrm{s}}} \right)}_{:,r}}},
	\label{eq:CPdecomposition}
\end{equation}
where, ${{\left( \mathbf{A} \right)}_{:,r}},{{\left( \mathbf{A} \right)}_{:,r}},{{\left( {{\mathbf{R}}_{\mathrm{s}}} \right)}_{:,r}}$ are the $r$-th columns of each factor matrix, $\circ $ is outer product, and  $()_{:}$ defines elements of matrices. In CP decompositon, ${{{\bm{\mathcal{R}}}_{\mathbf{x}}}}$ is expressed as a summation of rank-one tensors. The Eq.(\ref{eq:CPdecomposition}) can be rewritten as:
\begin{equation}
	\begin{split}
		&{\tilde{\mathbf{{R}}}_{\mathbf{x}}}\left( {{\tau }_{1}} \right)=\mathbf{{A}}{\tilde{\mathbf{{R}}}_{\mathbf{s}}}\left( {{\tau }_{1}} \right){{\mathbf{{A}}}^{\mathrm{T}}}
		\ \mathrm{where}\ {\tilde{\mathbf{{R}}}_{\mathbf{s}}}\left( {{\tau }_{1}} \right) = 
		\mathrm{diag}({{\left( {{\mathbf{R}}_{\mathrm{s}}} \right)}_{1,:}})\\
		&\ \ \ \ \ \ \ \ \vdots\\
		&{\tilde{\mathbf{{R}}}_{\mathbf{x}}}\left( {{\tau }_{\kappa}} \right)=\mathbf{{A}}{\tilde{\mathbf{{R}}}_{\mathbf{s}}}\left( {{\tau }_{\kappa}} \right){{\mathbf{{A}}}^{\mathrm{T}}}
		\ \mathrm{where}\ {\tilde{\mathbf{{R}}}_{\mathbf{s}}}\left( {{\tau }_{\kappa}} \right) = 
		\mathrm{diag}({{\left( {{\mathbf{R}}_{\mathrm{s}}} \right)}_{\kappa,:}})\\
		&\ \ \ \ \ \ \ \ \vdots\\
		&{\tilde{\mathbf{{R}}}_{\mathbf{x}}}\left( {{\tau }_{K}} \right)=\mathbf{{A}}{\tilde{\mathbf{{R}}}_{\mathbf{s}}}\left( {{\tau }_{K}} \right){{\mathbf{{A}}}^{\mathrm{T}}}
		\ \mathrm{where}\ {\tilde{\mathbf{{R}}}_{\mathbf{s}}}\left( {{\tau }_{K}} \right) = 
		\mathrm{diag}({{\left( {{\mathbf{R}}_{\mathrm{s}}} \right)}_{K,:}}).\\
	\end{split}
	\label{eq:CPdecomposition_SOBI}
\end{equation}
Therefore, covariance matrices can be simultaneously diagonalized via CP decomposition and CP rank $D$ corresponds to the number of identified vibration modes $\mathbf{A}$.

As CP decomposition assures uniqueness of the decomposition when the lowest dimension of tensor (min($M,M,K$)) is lower than the CP rank $D$, BSS can be applied for identification of under-determined systems. The input signals that can be identified against the number of sensors are limited by the mathematical conditions to satisfy the uniqueness of CP decomposition by Kruskal (Kruskal, 1977):
%Kruscal condition
\begin{equation}
	{{k}_{\mathbf{A}}}+{{k}_{\mathbf{A}}}+{{k}_{\mathbf{R}}}\ge 2D+2,
	\label{eq:Kruscal condition}
\end{equation}
where, ${{k}_{\mathbf{A}}}$, ${{k}_{\mathbf{R}}}$ are ranks of $\mathbf{A}$ and $\mathbf{R}$, respectively. The condition can be relaxed between numbers of output and input signals $M$ and $N$ as:
\begin{equation}
	\frac{M\left( M-1 \right)}{2}\le \frac{N\left( N-1 \right)}{4}\left( \frac{N\left( N-1 \right)}{2}+1 \right)-\frac{N!}{\left( N-4 \right)!4!}{{c}_{\left\{ N\ge 4 \right\}}}\ \mathrm{where}\ {{c}_{\left\{ N\ge 4 \right\}}}:=\left\{ \begin{matrix}
		0,\ \mathrm{if}\ N<4  \\
		1,\ \mathrm{if}\ N\geq4  \\
	\end{matrix} \right.,\
	\label{eq:Sensor_limitation}
\end{equation}
(Stegeman et al., 2006). From eq. (\ref{eq:Sensor_limitation}), the maximum numbers of input signals $N$ that can be identified from $M$-output signals are determined, as shown in Table 1. Specifically, the ideal numbers that can be identified for the numbers of sensors can be determined.

\begin{table}[H]
	\begin{center}
		\caption{Relationship between the numbers of measured output signals and input signals which can be estimated.}
		\begin{tabular}{l|ccccccccccc}
			\hline
			Measured output signals($M$)&2&3&4&5&6&7&8&9&10&11&12\\
			\hline
			Input signals which can be estimated ($N$)&2&4&6&10&15&20&26&33&41&49&59\\
			\hline
		\end{tabular}
	\end{center}
	\label{table:Sensor_limitation}
\end{table}

\section{\label{proposal}Operational modal analysis for under-determined systems via Bayesian CP decomposition}
To automatically determine the numbers of vibration modes from under-determined systems, CP ranks are identified by applying BCPF (Zhao et al., 2015) for OMA, where probabilistic CP models are identified via variational Bayesian inference (Bishop, 2008) by considering three-dimensional tensors as probabilistic parameters. As a result of the CP decomposition, the mixture models (vibration modes) and input signals (modal responses) are identified.

\subsection{Probabilistic CP models and prior distribution}
This subsection provides joint distributions of the three-dimensional tensors constructed by stacking of covariance matrices shown in Eq. (\ref{eq:CPdecomposition_1}). First, probabilistic CP models with noise ${{{\bm{\mathcal{R}}}_{\mathbf{x}}}}$ are denoted as:
\begin{equation}
	p\left( \left. {{\bm{{\mathcal{R}}}}_{\mathbf{x}}} \right|\mathbf{A},\mathbf{A},{{\mathbf{R}}_{\mathbf{s}}},\beta  \right)=\prod\limits_{i=1}^{M}{\prod\limits_{j=1}^{M}{\prod\limits_{k=1}^{K}{\mathsf{\mathcal{N}}\left( \left. {{\left( {{\bm{{\mathcal{R}}}}_{\mathbf{x}}} \right)}_{ijk}} \right|\sum\limits_{r=1}^{D}{{{\left( \mathbf{A} \right)}_{:,r}}\circ {{\left( \mathbf{A} \right)}_{:,r}}\circ {{\left( {{\mathbf{R}}_{\mathrm{s}}} \right)}_{:,r}}},{{\beta }^{-1}} \right)}}},
\end{equation}
where, $\mathcal{N}(\mu,\sigma^2)$ is a Gaussian distribution with average $\mu$, variance $\sigma^2$, and $\beta$ is the noise precision.

Next, the prior distribution of the factor matrices for CP decomposition is given as:
\begin{align}
	& p\left( \left. \mathbf{A} \right|\mathbf{\lambda } \right)=\prod\limits_{i=1}^{M}{\mathsf{\mathcal{N}}\left( \left. {{\left( \mathbf{A} \right)}_{i,:}} \right|\mathbf{0},{{\mathbf{\Lambda }}^{-1}} \right)} \\ 
	& p\left( \left. \mathbf{A} \right|\mathbf{\lambda } \right)=\prod\limits_{j=1}^{M}{\mathsf{\mathcal{N}}\left( \left. {{\left( \mathbf{A} \right)}_{j,:}} \right|\mathbf{0},{{\mathbf{\Lambda }}^{-1}} \right)} \\ 
	& p\left( \left. {{\mathbf{R}}_{\mathrm{s}}} \right|\mathbf{\lambda } \right)=\prod\limits_{k=1}^{K}{\mathsf{\mathcal{N}}\left( \left. {{\left( {{\mathbf{R}}_{\mathrm{s}}} \right)}_{k,:}} \right|\mathbf{0},{{\mathbf{\Lambda }}^{-1}} \right)},
	\label{eq:P_factor}
\end{align}
where, $\mathbf{\Lambda }\in {{\mathbb{R}}^{D\times D}}$ is a diagonal matrix by the $r$-th factor $\bm{\lambda }\in {{\mathbb{R}}^{D}}$ and given as:
\begin{equation}
	\bm{\Lambda }=\mathrm{diag}\left( \bm{\lambda } \right),
	\label{eq:precision}
\end{equation}
which is also called the precision matrix. Next, the prior distributions of parameters $\bm{\lambda}$ are defined as:
\begin{equation}
	p\left( \bm{\lambda } \right)=\prod\limits_{r=1}^{D}{\mathrm{Ga}\left( \left. {{\lambda }_{r}} \right|c_{0}^{r},d_{0}^{r} \right)},
	\label{eq:prior_precision}
\end{equation}
where, $D$ is the maximum CP rank that we consider for CP decomposition, and Ga is a Gamma distribution defined as:
%\newpage
\begin{equation}
	\mathrm{Ga}\left( \left. x \right|a,b \right)=\frac{{{b}^{a}}{{x}^{a-1}}{{e}^{-bx}}}{\Gamma \left( a \right)},
	\label{eq:gamma_function}
\end{equation}
where, $\Gamma(x)$ is the Gamma function, $a$ and $b$ denote hyper parameters. In addition, the prior distribution of noise precision $\beta$ is also defined as:
\begin{equation}
	p\left( \beta  \right)=\mathrm{Ga}\left( \left. \beta  \right|a_{0}^{{}},b_{0}^{{}} \right).
	\label{eq:prior_beta}
\end{equation}
Therefore, for parameters $\mathbf{\Theta }=\left\{ \mathbf{A},\mathbf{A},{{\mathbf{R}}_{\mathrm{s}}},\bm{\lambda },\beta  \right\}$, the joint distribution of tensor ${\bm{{\mathcal{R}}}_{\mathbf{x}}}$ and ${\mathbf{\Theta}}$ is defined as:
\begin{equation}
	p\left( {\bm{{\mathcal{R}}}_{\mathbf{x}}},\mathbf{\Theta } \right)=p\left( \left. {\bm{{\mathcal{R}}}_{\mathbf{x}}} \right|\mathbf{A},\mathbf{A},{{\mathbf{R}}_{\mathbf{s}}},\bm{\lambda },\beta  \right)p\left( \left. \mathbf{A} \right|\bm{\lambda } \right)p\left( \left. \mathbf{A} \right|\bm{\lambda } \right)p\left( \left. {{\mathbf{R}}_\mathbf{s}} \right|\bm{\lambda } \right)p\left( \bm{\lambda } \right)p\left( \beta  \right).
	\label{eq:joint_distribution}
\end{equation}

\subsection{Bayesian CP decomposition}
Probabilistic CP models for CP decomposition are determined using the variational Bayesian frameworks, where  the tensor ${{\bm{\mathcal{R}}}_{\mathbf{x}}}$ is used as observed data via the Kullback–Leibler (KL) divergence between the posterior distribution $p\left( \left. \mathbf{\Theta } \right|{{{\bm{\mathcal{R}}}_{\mathbf{x}}}} \right)$ and the approximated posterior $q\left( \mathbf{\Theta } \right)$:
\begin{align}
	\mathrm{KL}\left( q\left( \mathbf{\Theta } \right),p\left( \mathbf{\Theta }|{{\bm{{\mathcal{R}}}}_{\mathbf{x}}} \right) \right) &= \ln \int{q\left( \mathbf{\Theta } \right)\frac{q\left( \mathbf{\Theta } \right)}{p\left( \mathbf{\Theta }|{{\bm{{\bm{{\mathcal{R}}}}}}_{\mathbf{x}}} \right)}d\mathbf{\Theta }} \nonumber \\
	&= \ln p\left( {\bm{{\mathcal{R}}}_{\mathbf{x}}} \right)-\mathsf{\mathcal{L}}\left( q \right)
	\label{eq:KL}
\end{align}
where
\begin{equation}
	\mathsf{\mathcal{L}}\left( q \right)=\int{q\left( \mathbf{\Theta } \right)}\ln \left\{ \frac{p\left( {{\bm{{\mathcal{R}}}}_{\mathbf{x}}},\mathbf{\Theta } \right)}{q\left( \mathbf{\Theta } \right)} \right\}d\mathbf{\Theta},
	\label{eq:lower}
\end{equation}
is used. $\ln p\left( {{\bm{\mathcal{R}}}_{\mathbf{x}}} \right)$ is the model evidence and marginalization of likelihood function of ${{\bm{\mathcal{R}}}_{\mathbf{x}}}$ by parameter $\bm{\Theta}$. As the KL divergence approaches zero when the distributions equalize, minimization of KL divergence leads to approximation of $q\left( \mathbf{\Theta } \right)$. To solve the optimization problem using KL divergence, a convex function $f(x)$ of stochastic variables $x$ can be obtained from Jensen's inequality as:
\begin{equation}
	f(\mathbb{E}[x])\ge \mathbb{E}[f(x)],
	\label{eq:Jensen}
\end{equation}
which provides that $\mathsf{\mathcal{L}}\left( q \right)$ of Eq. (\ref{eq:KL}) is:
\begin{align}
	\mathsf{\mathcal{L}}\left( q \right) &\le \ln \int{q\left( \mathbf{\Theta } \right)\frac{q\left( \mathbf{\Theta } \right)}{p\left( \mathbf{\Theta }|{{\bm{\mathcal{R}}}_{\mathbf{x}}} \right)}d\mathbf{\Theta }} \nonumber \\
	&=\ln \int{p\left( {{\bm{\mathcal{R}}}_{\mathbf{x}}},\mathbf{\Theta } \right)d\mathbf{\Theta }}=\ln p\left( {{\bm{\mathcal{R}}}_{\mathbf{x}}} \right).
	\label{eq:Jemsem2}
\end{align}
Therefore, $\mathsf{\mathcal{L}}\left( q \right)$ is the lower bound of model evidence $\ln p\left( {\bm{{\mathcal{R}}}_{\mathbf{x}}} \right)$ shown in Eq. (\ref{eq:KL}), the KL divergence become zero when the distributions are equal. Specifically, the KL divergence is minimized by maximizing $\mathsf{\mathcal{L}}\left( q \right)$, which leads to $q\left( \mathbf{\Theta } \right)=p\left( \left. \mathbf{\Theta } \right|{\bm{{\mathcal{R}}}_{\mathbf{x}}} \right)$.

For variational Bayesian frameworks, assuming that $\bm{\Theta}_j$ are independent, $q\left( \mathbf{\Theta } \right)$ is approximated as:
\begin{equation}
	q\left( \mathbf{\Theta } \right)=\prod\limits_{j=1}^{{}}{{{q}_{{{\mathbf{\Theta }}_{j}}}}\left( {{\mathbf{\Theta }}_{j}} \right)},
	\label{eq:Joint}
\end{equation}
where, ${{q}_{\mathbf{\Theta }}}\left( \mathbf{\Theta } \right)=\left\{ {{q}_{\mathbf{A}}}\left( \mathbf{A} \right),{{q}_{\mathbf{A}}}\left( \mathbf{A} \right),{{q}_{{{\mathbf{R}}_{\mathrm{s}}}}}\left( {{\mathbf{R}}_{\mathrm{s}}} \right),{{q}_{\bm{\lambda }}}\left( \bm{\lambda } \right),{{q}_{\beta }}\left( \beta  \right) \right\}$. Optimal ${{q}_{{{\mathbf{\Theta }}_{j}}}}\left( {{\mathbf{\Theta }}_{j}} \right)$ is given as:
\begin{equation}
	\ln {{q}_{{{\mathbf{\Theta }}_{j}}}}\left( {{\mathbf{\Theta }}_{j}} \right)={{\mathbb{E}}_{q\left( \mathbf{\Theta }/{{\mathbf{\Theta }}_{j}} \right)}}\left[ \ln p\left( {\bm{{\mathcal{R}}}_{\mathbf{x}}},\mathbf{\Theta } \right) \right]+\mathrm{const}.,
	\label{eq:Optimum}
\end{equation}
where, ${{\mathbb{E}}_{q\left( \mathbf{\Theta }/{{\mathbf{\Theta }}_{j}} \right)}}\left[ \cdot  \right]$ is the exception of posterior distribution except for $j$-th factors $\bm{\Theta}$. In \ref{subsub:factor} -- \ref{subsub:noise}, each posterior distribution is shown and the model evidence is defined in \ref{subsub:lower}. Finally, determination of numbers of input signals are described in \ref{subsub:CP rank}.
\vspace{0.5cm}

\subsubsection{\label{subsub:factor}Posterior distribution of factor matrices}
From Eq. (\ref{eq:Optimum}), the posterior distributions of factor matrices of CP decomposition are expressed by:
\begin{align}
	& {{q}_{\mathbf{A}}}\left( \mathbf{A} \right)=\prod\limits_{i=1}^{M}{\mathsf{\mathcal{N}}\left( \left. {{\left( \mathbf{A} \right)}_{i,:}} \right|\left( {\mathbf{\tilde{A}}} \right)_{i,:}^{{}},\mathbf{V}_{i}^{\left( \mathbf{A} \right)} \right)} \\ 
	& {{q}_{\mathbf{A}}}\left( \mathbf{A} \right)=\prod\limits_{j=1}^{M}{\mathsf{\mathcal{N}}\left( \left. {{\left( \mathbf{A} \right)}_{j,:}} \right|\left( {\mathbf{\tilde{A}}} \right)_{j,:}^{{}},\mathbf{V}_{j}^{\left( \mathbf{A} \right)} \right)} \\ 
	& {{q}_{{{\mathbf{R}}_{\mathrm{s}}}}}\left( {{\mathbf{R}}_{\mathrm{s}}} \right)=\prod\limits_{k=1}^{K}{\mathsf{\mathcal{N}}\left( \left. {{\left( {{\mathbf{R}}_{\mathrm{s}}} \right)}_{k,:}} \right|\left( {{{\mathbf{\tilde{R}}}}_{\mathrm{s}}} \right)_{k,:}^{{}},\mathbf{V}_{k}^{\left( {{\mathbf{R}}_{\mathrm{s}}} \right)} \right)}
	\label{eq:CP post}
\end{align}
using Gamma distribution, where average of posterior distribution $\mathbf{\tilde{A}}\in {{\mathbb{R}}^{M\times D}},{{\mathbf{\tilde{R}}}_{s}}\in {{\mathbb{R}}^{K\times D}}$ and variance $\mathbf{V}_{{}}^{\left( \mathbf{A} \right)}\in {{\mathbb{R}}^{D\times D}},\mathbf{V}_{{}}^{\left( {{\mathbf{R}}_{\mathrm{s}}} \right)}\in {{\mathbb{R}}^{D\times D}}$ are:
\begin{align}
	& \left( {\mathbf{\tilde{A}}} \right)_{i,:}^{{}}={{\mathbb{E}}_{q}}\left[ \beta  \right]\mathbf{V}_{i}^{\left( \mathbf{A} \right)}{{\mathbb{E}}_{q}}\left[ {{\left( {{\mathbf{R}}_{\mathrm{s}}}\odot \mathbf{A} \right)}_{i}^{\mathrm{T}}} \right]\mathrm{vec}\left( {{\left( {{\bm{\mathcal{R}}}_{\mathbf{x}}} \right)}_{i,:,:}}, \right)\nonumber\\ 
	& \ \ \ \ \ \ \ \ ={{\mathbb{E}}_{q}}\left[ \beta  \right]\mathbf{V}_{i}^{\left( \mathbf{A} \right)}{{\left( {{\mathbb{E}}_{q}}\left[ {{\mathbf{R}}_{\mathrm{s}}} \right]\odot {{\mathbb{E}}_{q}}\left[ \mathbf{A} \right] \right)}^{\mathrm{T}}}\mathrm{vec}\left( {{\left( {{\bm{\mathcal{R}}}_{\mathbf{x}}} \right)}_{i,:,:}} \right),\\ 
	& \left( {\mathbf{\tilde{A}}} \right)_{j,:}^{{}}={{\mathbb{E}}_{q}}\left[ \beta  \right]\mathbf{V}_{j}^{\left( \mathbf{A} \right)}{{\mathbb{E}}_{q}}\left[ {{\left( {{\mathbf{R}}_{\mathrm{s}}}\odot \mathbf{A} \right)}_{j}^{\mathrm{T}}} \right]\mathrm{vec}\left( {{\left( {{\bm{\mathcal{R}}}_{\mathbf{x}}} \right)}_{:,j,:}} \right)\nonumber\\ 
	& \ \ \ \ \ \ \ \ \ ={{\mathbb{E}}_{q}}\left[ \beta  \right]\mathbf{V}_{j}^{\left( \mathbf{A} \right)}{{\left( {{\mathbb{E}}_{q}}\left[ {{\mathbf{R}}_{\mathrm{s}}} \right]\odot {{\mathbb{E}}_{q}}\left[ \mathbf{A} \right] \right)}^{\mathrm{T}}}\mathrm{vec}\left( {{\left( {{\bm{\mathcal{R}}}_{\mathbf{x}}} \right)}_{:,j,:}} \right) ,\\ 
	& \left( {{{\mathbf{\tilde{R}}}}_{s}} \right)_{k,:}^{{}}={{\mathbb{E}}_{q}}\left[ \beta  \right]\mathbf{V}_{k}^{\left( {{\mathbf{R}}_{\mathrm{s}}} \right)}{{\mathbb{E}}_{q}}\left[ {{\left( \mathbf{A}\odot \mathbf{A} \right)}_{k}^{\mathrm{T}}} \right]\mathrm{vec}\left( {{\left( {{\bm{\mathcal{R}}}_{\mathbf{x}}} \right)}_{:,:,k}} \right)\nonumber \\ 
	& \ \ \ \ \ \ \ \ \ \ ={{\mathbb{E}}_{q}}\left[ \beta  \right]\mathbf{V}_{k}^{\left( {{\mathbf{R}}_{\mathrm{s}}} \right)}{{\left( {{\mathbb{E}}_{q}}\left[ \mathbf{A} \right]\odot {{\mathbb{E}}_{q}}\left[ \mathbf{A} \right] \right)}^{\mathrm{T}}}\mathrm{vec}\left( {{\left( {{\bm{\mathcal{R}}}_{\mathbf{x}}} \right)}_{:,:,k}} \right),
\end{align}
\begin{align}
	& \mathbf{V}_{i}^{\left( \mathbf{A} \right)}={{\left( {{\mathbb{E}}_{q}}\left[ \beta  \right]{{\mathbb{E}}_{q}}\left[ {{\left( {{\mathbf{R}}_{\mathrm{s}}}\odot \mathbf{A} \right)}_{i}^{\mathrm{T}}}{{\left( {{\mathbf{R}}_{\mathrm{s}}}\odot \mathbf{A} \right)}_{i}} \right]+{{\mathbb{E}}_{q}}\left[ \mathbf{\Lambda } \right] \right)}^{-1}} ,\\ 
	& \mathbf{V}_{j\,\,\,}^{\left( \mathbf{A} \right)}={{\left( {{\mathbb{E}}_{q}}\left[ \beta  \right]{{\mathbb{E}}_{q}}\left[ {{\left( {{\mathbf{R}}_{\mathrm{s}}}\odot \mathbf{A} \right)}_{j}^{\mathrm{T}}}{{\left( {{\mathbf{R}}_{\mathrm{s}}}\odot \mathbf{A} \right)}_{j}} \right]+{{\mathbb{E}}_{q}}\left[ \mathbf{\Lambda } \right] \right)}^{-1}} ,\\ 
	& \mathbf{V}_{k}^{\left( {{\mathbf{R}}_{\mathrm{s}}} \right)}={{\left( {{\mathbb{E}}_{q}}\left[ \beta  \right]{{\mathbb{E}}_{q}}\left[ {{\left( \mathbf{A}\odot \mathbf{A} \right)}_{k}^{\mathrm{T}}}{{\left( \mathbf{A}\odot \mathbf{A} \right)}_{k}} \right]+{{\mathbb{E}}_{q}}\left[ \mathbf{\Lambda } \right] \right)}^{-1}},
\end{align}
where, $\odot $ is the Khatori-Rao product.

\vspace{0.5cm}
\subsubsection{\label{subsub:seido}Posterior distribution of precision parameters }
Using observed data on $L$-points, from Eq. (\ref{eq:Optimum}), the posterior distribution of precision matrix of factor matrix $\bm{\lambda }$ is expressed by Gamma distribution as:
\begin{equation}
	{{q}_{\bm{\lambda }}}\left( \bm{\lambda } \right)=\prod\limits_{r=1}^{D}{\mathrm{Ga}\left( \left. {{\lambda }_{r}} \right|c_{L}^{r},d_{L}^{r} \right)},
	\label{eq:q lambda}
\end{equation}
where, hyper parameters are defined as:
\begin{align}
	& c_{L}^{r}=c_{0}^{r}+\frac{2M+K}{2}, \\ 
	& d_{L}^{r}=d_{0}^{r}+\frac{1}{2}\left( {{\mathbb{E}}_{q}}\left[ \left( \mathbf{A} \right)_{:,r}^{\mathrm{T}}{{\left( \mathbf{A} \right)}_{:,r}} \right]+{{\mathbb{E}}_{q}}\left[ \left( \mathbf{A} \right)_{:,r}^{\mathrm{T}}{{\left( \mathbf{A} \right)}_{:,r}} \right]+{{\mathbb{E}}_{q}}\left[ \left( {{\mathbf{R}}_{\mathrm{s}}} \right)_{:,r}^{\mathrm{T}}{{\left( {{\mathbf{R}}_{\mathrm{s}}} \right)}_{:,r}} \right] \right),
\end{align}
where, exception of inner products of the $r$-th factor in second equation is 
\begin{align}
	& {{\mathbb{E}}_{q}}\left[ \left( \mathbf{A} \right)_{:,r}^{\mathrm{T}}{{\left( \mathbf{A} \right)}_{:,r}} \right]=\left( {\mathbf{\tilde{A}}} \right)_{:,r}^{\mathrm{T}}{{\left( {\mathbf{\tilde{A}}} \right)}_{:,r}}+{{\sum\limits_{i}{\left( \mathbf{V}_{i}^{\left( \mathbf{A} \right)} \right)_{r,r}}}}, \\ 
	& {{\mathbb{E}}_{q}}\left[ \left( \mathbf{A} \right)_{:,r}^{\mathrm{T}}{{\left( \mathbf{A} \right)}_{:,r}} \right]=\left( {\mathbf{\tilde{A}}} \right)_{:,r}^{\mathrm{T}}{{\left( {\mathbf{\tilde{A}}} \right)}_{:,r}}+{{\sum\limits_{j}{\left( \mathbf{V}_{j}^{\left( \mathbf{A} \right)} \right)_{r,r}}}} ,\\ 
	& {{\mathbb{E}}_{q}}\left[ \left( {{\mathbf{R}}_{\mathrm{s}}} \right)_{:,r}^{\mathrm{T}}{{\left( {{\mathbf{R}}_{\mathrm{s}}} \right)}_{:,r}} \right]=\left( {{{\mathbf{\tilde{R}}}}_{\mathrm{s}}} \right)_{:,r}^{\mathrm{T}}{{\left( {{{\mathbf{\tilde{R}}}}_{\mathrm{s}}} \right)}_{:,r}}+{{\sum\limits_{k}{\left( \mathbf{V}_{k}^{\left( {{\mathbf{R}}_{\mathrm{s}}} \right)} \right)_{r,r}}}},
\end{align}
where, ${\lambda}_{r}$ of Eq. (\ref{eq:q lambda})is updated by ${L}_{2}$ norm of $r$-th factor, which means the exception of ${\lambda}_{r}$ becomes larger as the $L_2$ norm decreases. Therefore, factors with small precision converge to zero as a result of updating.

\vspace{0.5cm}
\subsubsection{\label{subsub:noise}Posterior distribution of noise precision $\beta$}
From Eq. (\ref{eq:Optimum}), the posterior distribution of the noise precision $\beta$ can be expressed by a Gamma distribution as: 
\begin{equation}
	{{q}_{\beta }}\left( \beta  \right)=\mathrm{Ga}\left( \left. \beta  \right|a_{L}^{{}},b_{L}^{{}} \right),
	\label{eq:q beta}
\end{equation}
where, the posterior distribution ${{q}_{\beta }}\left( \beta  \right)$ is defined by hyper parameters:
\begin{align}
	& a_{L}^{{}}=a_{0}^{{}}+\frac{1}{2}{{M}^{2}}K \nonumber, \\ 
	& b_{L}^{{}}=b_{0}^{{}}+\frac{1}{2}{{\mathbb{E}}_{q}}\left[ \left\| {\bm{{\mathcal{R}}}_{\mathbf{x}}}-\sum\limits_{r=1}^{D}{{{\left( \mathbf{A} \right)}_{:,r}}\circ {{\left( \mathbf{A} \right)}_{;,r}}\circ {{\left( {{\mathbf{R}}_{\mathrm{s}}} \right)}_{:,r}}} \right\|_{F}^{2} \right],
	\label{eq:hyper beta}
\end{align}
where, the exception of the second term in Eq. (\ref{eq:hyper beta}) is:
\begin{align}
	& {{\mathbb{E}}_{q}}\left[ \left\| {{\bm{\mathcal{R}}}_{\mathbf{x}}}-\sum\limits_{r=1}^{D}{{{\left( \mathbf{A} \right)}_{:,r}}\circ {{\left( \mathbf{A} \right)}_{;,r}}\circ {{\left( {{\bm{\mathbf{R}}}_{\mathrm{s}}} \right)}_{:,r}}} \right\|_{F}^{2} \right] \nonumber \\ 
	& =\left\| {{\bm{\mathcal{R}}}_{\mathbf{x}}} \right\|_{F}^{2}-2{{\mathrm{vec}}^{\mathrm{T}}}\left( {{\bm{\mathcal{R}}}_{\mathbf{x}}} \right)\mathrm{vec}\left( \sum\limits_{r=1}^{D}{{{\left( \mathbf{A} \right)}_{:,r}}\circ {{\left( \mathbf{A} \right)}_{;,r}}\circ {{\left( {{\bm{\mathbf{R}}}_{\mathrm{s}}} \right)}_{:,r}}} \right)+\left\| \mathbf{B}_{{}}^{\left( {{\mathbf{R}}_{\mathrm{s}}} \right)}\odot \mathbf{B}_{{}}^{\left( \mathbf{A} \right)}\odot \mathbf{B}_{{}}^{\left( \mathbf{A} \right)} \right\|,
	\label{Modelling error}
\end{align}
In addition, matrix  $\mathbf{B}\in {{\mathbb{R}}^{D\times D}}$ is expressed by quadratic forms of CP decomposition:
\begin{align}
	& {{\left( \mathbf{B}_{{}}^{\left( \mathbf{A} \right)} \right)}_{i,:}}=\mathrm{vec}\left( {{\mathbb{E}}_{q}}\left[ {{\left( \mathbf{A} \right)}_{i,:}}{{\left( \mathbf{A} \right)}_{i,:}}^{\mathrm{T}} \right] \right)=\mathrm{vec}\left( {{\left( {\mathbf{\tilde{A}}} \right)}_{i,:}}\left( {\mathbf{\tilde{A}}} \right)_{i,:}^{\mathrm{T}}+\mathbf{V}_{i}^{\left( \mathbf{A} \right)} \right) ,\\ 
	& {{\left( \mathbf{B}_{{}}^{\left( \mathbf{A} \right)} \right)}_{j,:}}=\mathrm{vec}\left( {{\mathbb{E}}_{q}}\left[ {{\left( \mathbf{A} \right)}_{j,:}}{{\left( \mathbf{A} \right)}_{j,:}}^{\mathrm{T}} \right] \right)=\mathrm{vec}\left( {{\left( {\mathbf{\tilde{A}}} \right)}_{j,:}}\left( {\mathbf{\tilde{A}}} \right)_{j,:}^{\mathrm{T}}+\mathbf{V}_{j}^{\left( \mathbf{A} \right)} \right) ,\\ 
	& {{\left( \mathbf{B}_{{}}^{\left( {{\mathbf{R}}_{\mathrm{s}}} \right)} \right)}_{k,:}}=\mathrm{vec}\left( {{\mathbb{E}}_{q}}\left[ {{\left( {{\mathbf{R}}_{\mathrm{s}}} \right)}_{k,:}}{{\left( {{\mathbf{R}}_{\mathrm{s}}} \right)}_{k,:}}^{\mathrm{T}} \right] \right)=\mathrm{vec}\left( {{\left( {{{\mathbf{\tilde{R}}}}_{\mathrm{s}}} \right)}_{k,:}}\left( {{{\mathbf{\tilde{R}}}}_{\mathrm{s}}} \right)_{k,:}^{\mathrm{T}}+\mathbf{V}_{k}^{\left( {{\mathbf{R}}_{\mathrm{s}}} \right)} \right).
\end{align}

\vspace{0.5cm}
\subsubsection{\label{subsub:lower}Lower bound of model evidence}
From Eq. (\ref{eq:CP post}), Eq. (\ref{eq:q lambda}) and Eq.(\ref{eq:q beta}), the lower bound of model evidence shown in Eq. (\ref{eq:lower}) is:
\begin{align}
	& \mathsf{\mathcal{L}}\left( q \right)=-\frac{{{a}_{L}}}{2{{b}_{L}}}{{\mathbb{E}}_{q}}\left[ \left\| {\bm{{\mathcal{R}}}_{\mathbf{x}}}-\sum\limits_{r=1}^{D}{{{\left( \mathbf{A} \right)}_{:,r}}\circ {{\left( \mathbf{A} \right)}_{;,r}}\circ {{\left( {{\mathbf{R}}_{\mathrm{s}}} \right)}_{:,r}}} \right\|_{F}^{2} \right] \nonumber \\ 
	& \ \ \ \ \ \ \ \ \ \ \ -\frac{1}{2}\mathrm{Tr}\left\{ \mathbf{\tilde{\Lambda }}\left( {{{\mathbf{\tilde{A}}}}^{\mathrm{T}}}\mathbf{\tilde{A}}+\sum\limits_{i}{\mathbf{V}_{i}^{\left( \mathbf{A} \right)}+{{{\mathbf{\tilde{A}}}}^{\mathrm{T}}}\mathbf{\tilde{A}}+\sum\limits_{j}{\mathbf{V}_{j}^{\left( \mathbf{A} \right)}}+{{{\mathbf{\tilde{R}}}}_{\mathrm{s}}}^{\mathrm{T}}{{{\mathbf{\tilde{R}}}}_{\mathrm{s}}}+\sum\limits_{k}{\mathbf{V}_{k}^{\left( {{\mathbf{R}}_{\mathrm{s}}} \right)}}} \right) \right\} \nonumber \\ 
	& \ \ \ \ \ \ \ \ \ \ \ +\frac{1}{2}\left( \sum\limits_{i}{\ln \left| \mathbf{V}_{i}^{\left( \mathbf{A} \right)} \right|+\sum\limits_{j}{\ln \left| \mathbf{V}_{j}^{\left( \mathbf{A} \right)} \right|}+\sum\limits_{k}{\ln \left| \mathbf{V}_{k}^{\left( {{\mathbf{R}}_{\mathrm{s}}} \right)} \right|}} \right)+\sum\limits_{r}{\left\{ \ln \Gamma \left( c_{L}^{r} \right) \right\}} \nonumber \\ 
	& \ \ \ \ \ \ \ \ \ \ \ +\sum\limits_{r}{\left\{ c_{L}^{r}\left( 1-\ln d_{L}^{r}-\frac{d_{0}^{r}}{d_{L}^{r}} \right)\  \right\}}\ +\ln \ \Gamma \left( {{a}_{L}} \right) \nonumber \\ 
	& \ \ \ \ \ \ \ \ \ \ \ +{{a}_{L}}\left( 1-\ln {{b}_{L}}-\frac{{{b}_{0}}}{{{b}_{L}}} \right)\ +\mathrm{const}.
	\label{eq:lower tensor}
\end{align}

\subsubsection{\label{subsub:CP rank}Automatic determination of vibration mode number}
During updating the posterior distribution of Eq. (\ref{eq:lower tensor}) described in the previous subsection, the factor matrices of CP decomposition $\mathbf{A}$ and $\mathbf{R_s}$ are estimated. In this process, as the lager precision parameters of the posterior distribution in Eq. (\ref{eq:precision}) $\lambda_r$ makes the $r$-th element to zero, the CP rank is automatically determined by the number of nonzero element updated by Bayesian inference, which delivers the automatic identification of the number of input signals to mixture models from only output signals.

\section{\label{validation}Numerical validation}
Using toy problems, we demonstrate that the true modal parameters are properly identified from under-determined systems via the proposed method.

\subsection{Toy problem}
\begin{figure}[H]
	\centering
	\includegraphics[clip, width=0.6\linewidth]{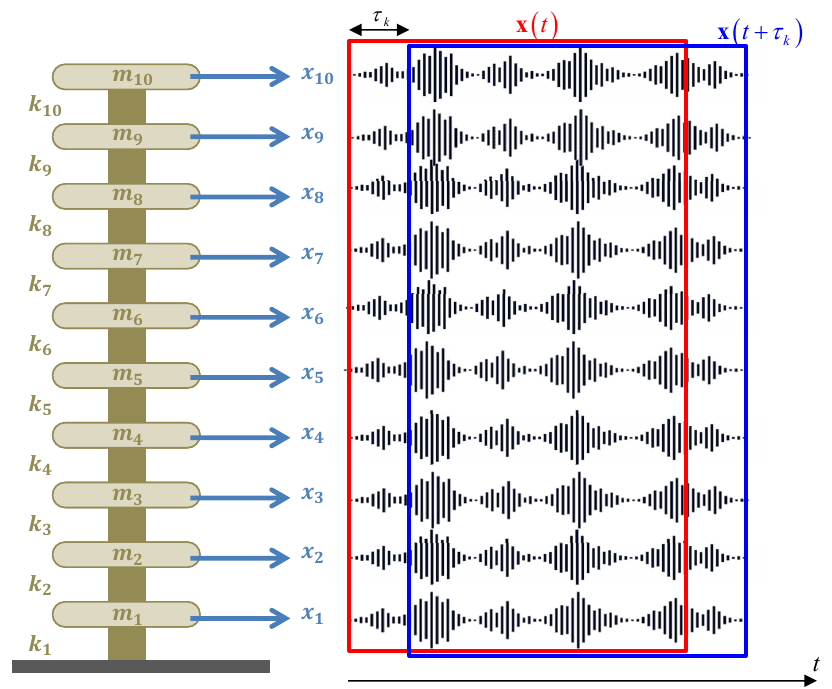}
	\caption{Measurement locations to generate artificial vibration data and the time delay of signals to calculate the covariance matrix.}
	\label{fig:toy problem}
\end{figure}

As a toy problem for OMA, we used mass-spring systems with mass and stiffness constants $m_i=100\ \mathrm{mt}\ (i=1,2,...,10)$ and $k_i=176.729\ \mathrm{MN/m}\ (i=1,2,...,10)$ (Abazarsa et al., 2013). In addition to the uniform mass and spring constants, non-uniform cases are tested by distributing mass and spring constants in the range $m_i\in[100, 200]\ \mathrm{mt}\ (i=1,2,...,10)$ and $k_i\in[176.729, 353.458]\ \mathrm{MN/m}(i=1,2,...,10)$. For these two mass-spring systems, the output signals $\bm{x}(t)$ of 10-degree-of-freedom (DOF)  are calculated in 180-s time with 100-Hz sampling frequency under random excitation to all DOFs as shown in Fig. \ref{fig:toy problem}. From the output signals obtained from toy problems, the natural frequencies, damping ratios, and modal vectors are estimated.

In the uniform and non-uniform mass and spring constants, the ground truth of natural frequencies and damping ratios are shown in Table 2, which are compared with results identified by OMA. In addition, we employ the modal assurance criterion (MAC) between analytical solutions and identified vectors:
\begin{equation}
	{\mathbf{MAC}\left(\bm{\phi}_a,\bm{\phi}_e\right)}=\frac{\left|{\bm{\phi}_a}^{\mathrm{T}}\bm{\phi}_e\right|}{\left({\bm{\phi}_a}^{\mathrm{T}}\bm{\phi}_a\right)\left({\bm{\phi}_e}^{\mathrm{T}}\bm{\phi}_e\right)}
	\label{eq:MAC}
\end{equation}
where ${\phi}_a$ and ${\phi}_e$ are analytical and identified modal vectors whose sizes correspond to the number of sensors.

\begin{figure}[H]
	\begin{center}
		\begin{tabular}{c}
			\begin{minipage}{0.2\hsize}
				\begin{center}
					\includegraphics[clip, width=1\linewidth]{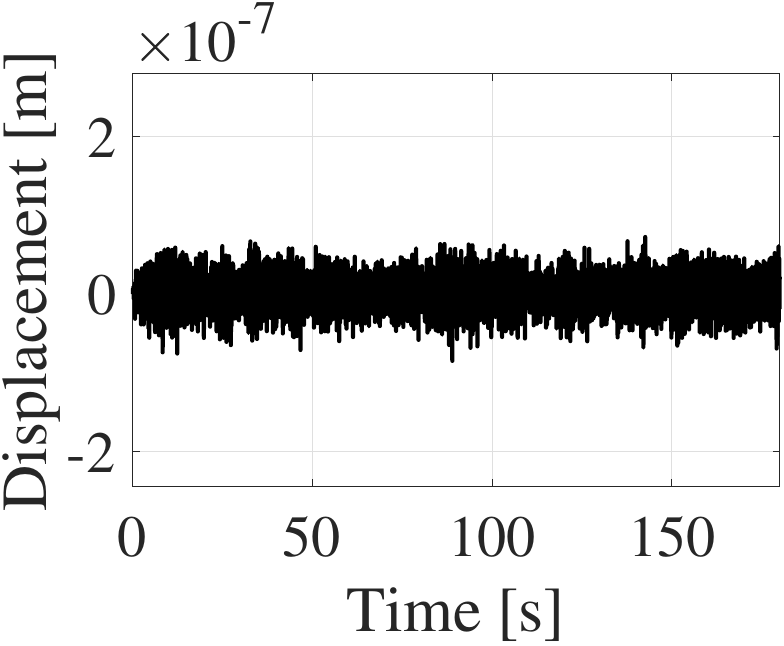}
					\hspace{1.6cm} (a) $x_1$
				\end{center}
			\end{minipage}
			
			% 2枚目の画像
			%			\hspace{0.5cm}
			\begin{minipage}{0.2\hsize}
				\begin{center}
					\includegraphics[clip, width=1\linewidth]{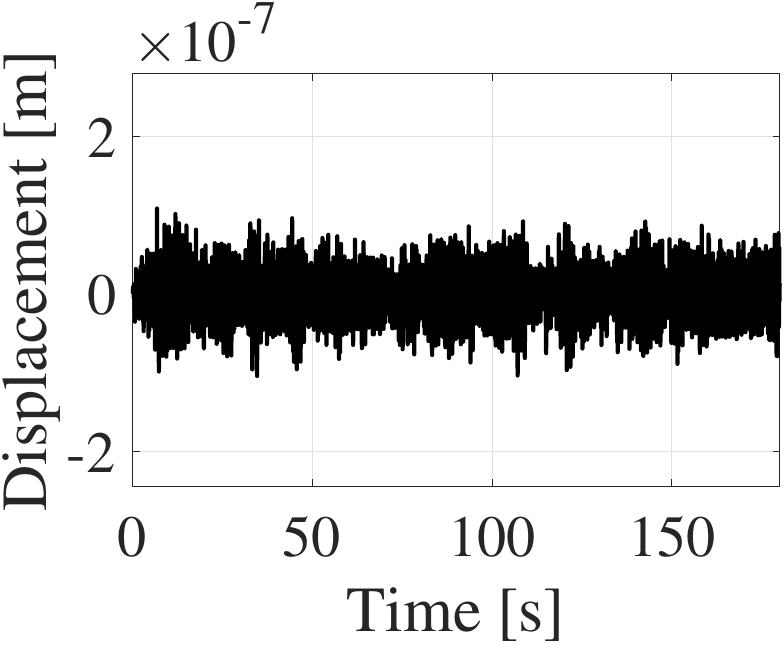}
					\hspace{1.6cm} (b) $x_2$
				\end{center}
			\end{minipage}
			
			%			\hspace{0.5cm}
			\begin{minipage}{0.2\hsize}
				\begin{center}
					\includegraphics[clip, width=1\linewidth]{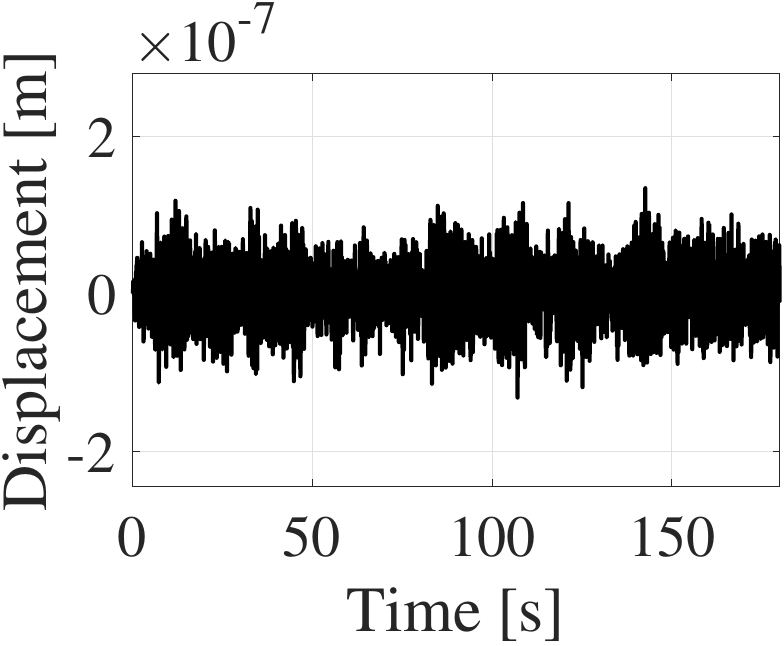}
					\hspace{1.6cm} (c) $x_3$
				\end{center}
			\end{minipage}
			
			%			\hspace{0.5cm}
			\begin{minipage}{0.2\hsize}
				\begin{center}
					\includegraphics[clip, width=1\linewidth]{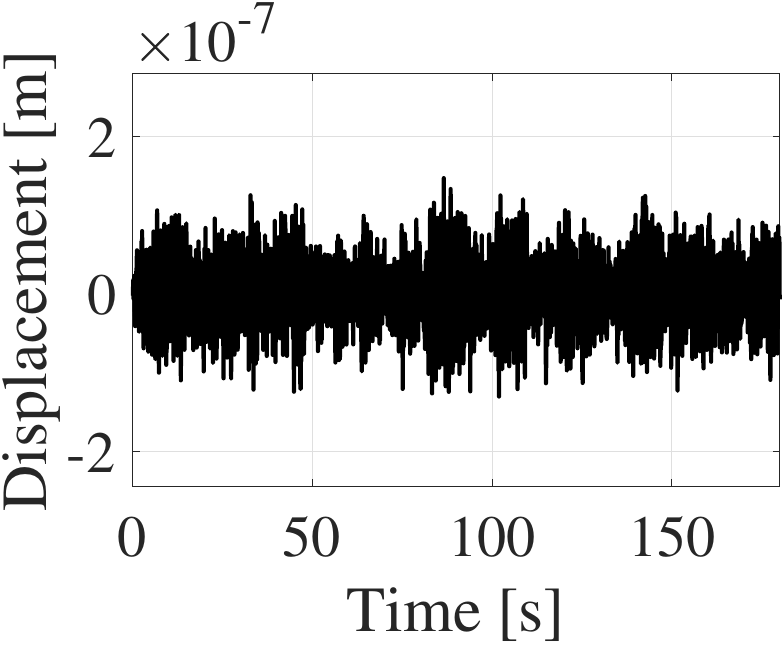}
					\hspace{1.6cm} (d) $x_4$
				\end{center}
			\end{minipage}
			
			%			\hspace{0.5cm}
			\begin{minipage}{0.2\hsize}
				\begin{center}
					\includegraphics[clip, width=1\linewidth]{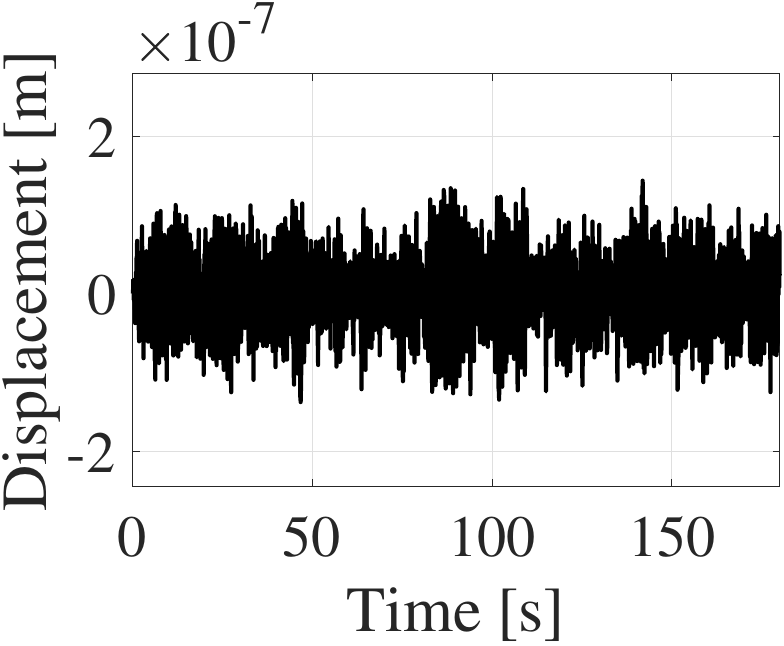}
					\hspace{1.6cm} (e) $x_5$
				\end{center}
			\end{minipage}
			
		\end{tabular}
	\end{center}		
	
	\vspace{0.2cm}
	\begin{center}
		\begin{tabular}{c}
			% 1枚目の画像
			\begin{minipage}{0.2\hsize}
				\begin{center}
					\includegraphics[clip, width=1\linewidth]{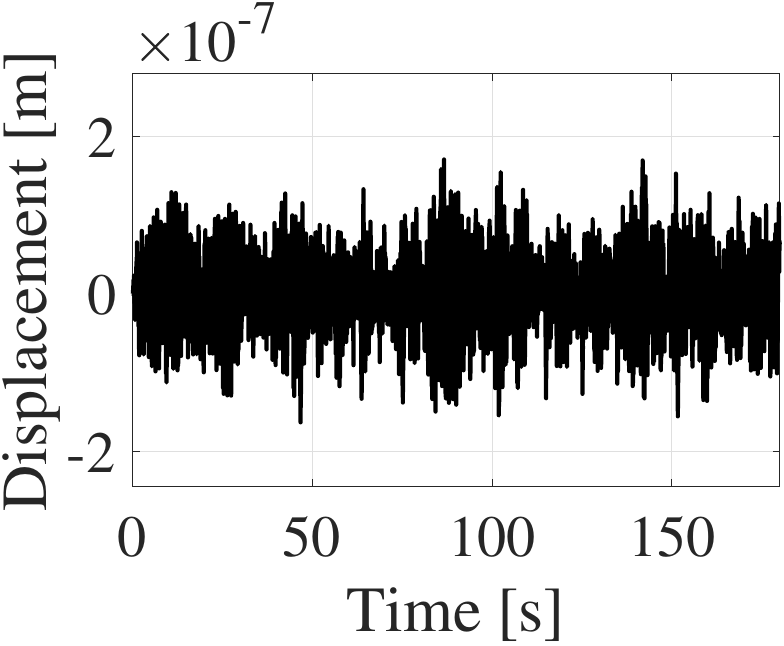}
					\hspace{1.6cm}  (f) $x_6$
				\end{center}
			\end{minipage}
			
			% 2枚目の画像
			%			\hspace{0.5cm}
			\begin{minipage}{0.2\hsize}
				\begin{center}
					\includegraphics[clip, width=1\linewidth]{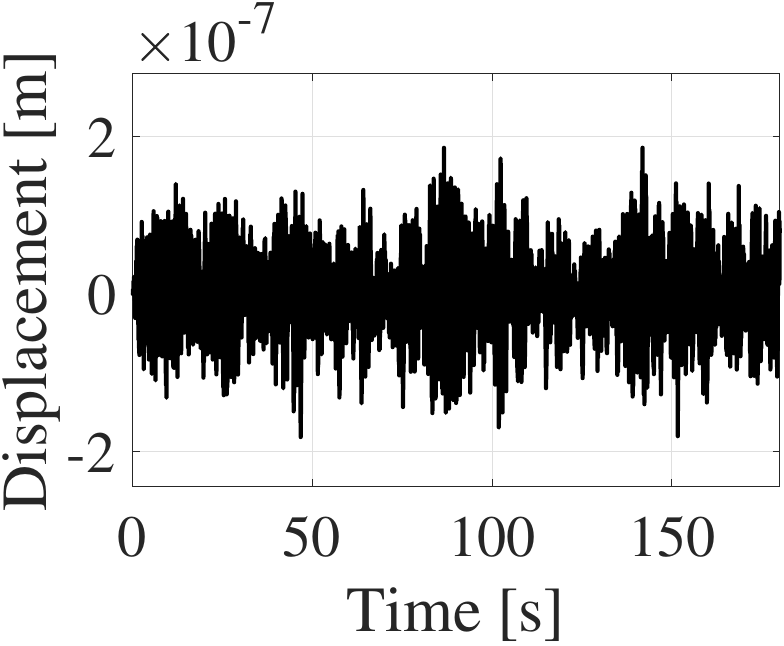}
					\hspace{1.6cm} (g) $x_7$
				\end{center}
			\end{minipage}
			
			%			\hspace{0.5cm}
			\begin{minipage}{0.2\hsize}
				\begin{center}
					\includegraphics[clip, width=1\linewidth]{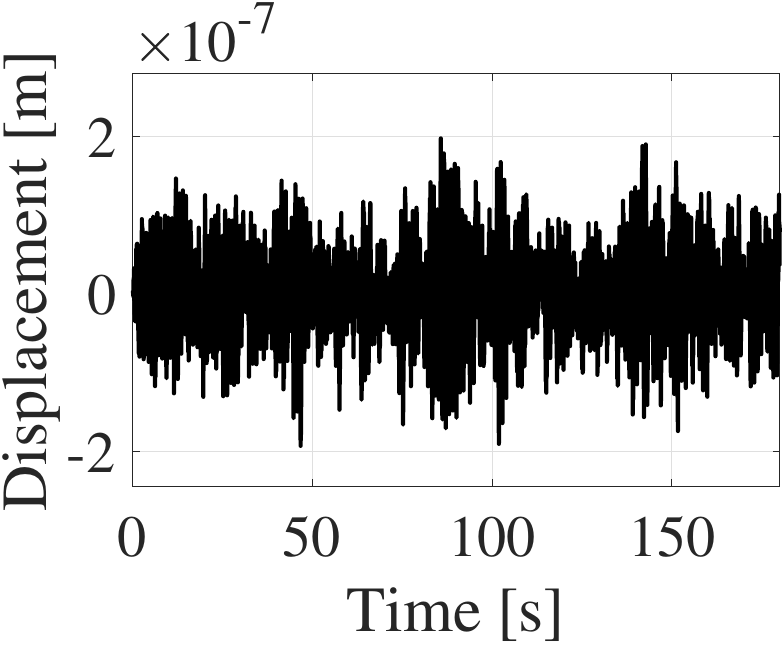}
					\hspace{1.6cm} (h) $x_8$
				\end{center}
			\end{minipage}
			
			%			\hspace{0.5cm}
			\begin{minipage}{0.2\hsize}
				\begin{center}
					\includegraphics[clip, width=1\linewidth]{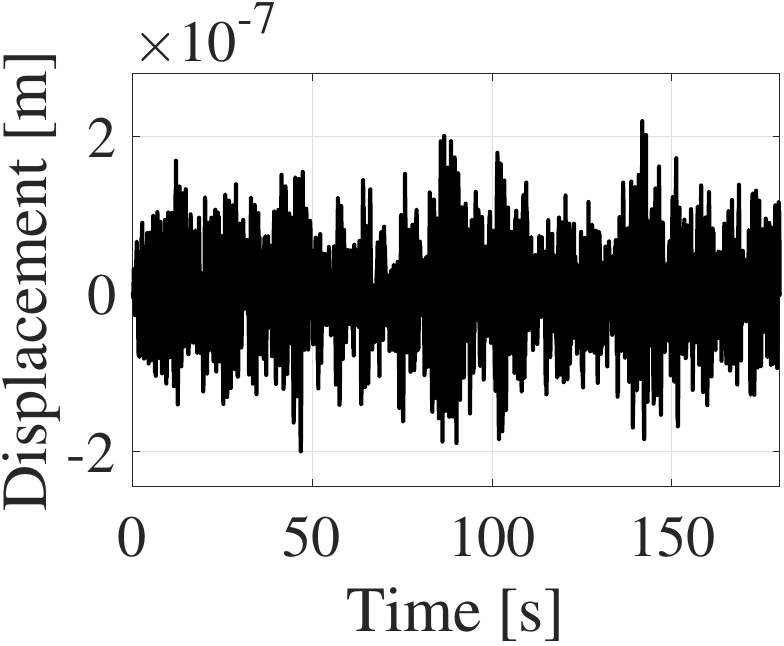}
					\hspace{1.6cm} (i) $x_9$
				\end{center}
			\end{minipage}
			
			%			\hspace{0.5cm}
			\begin{minipage}{0.2\hsize}
				\begin{center}
					\includegraphics[clip, width=1\linewidth]{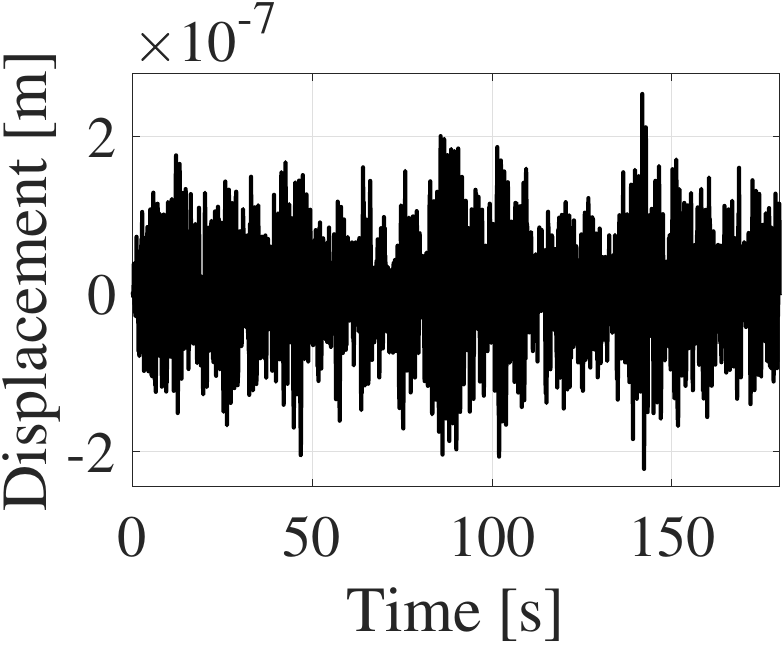}
					\hspace{1.6cm} (j) $x_{10}$
				\end{center}
			\end{minipage}
			
		\end{tabular}
	\end{center}		
	
	\caption{Examples of responses obtained by the 10-degree of freedom system.}
	\label{fig:Example of singnals}
\end{figure}

To demonstrate the identification for under-determined situations, OMA are performed using 5-10 sensors. Since these numbers of sensors satisfy the uniqueness of CP decomposition shown in Table 1, all modal parameters can be theoretically identified. To cover all combinations of sensor locations used for OMA, $M$-output signals are randomly extracted from the numerical simulations from mass $m_1,m_2,...,m_{10}$ for numbers of sensors $M=5,6,...,10$. Using output signals shown in Fig. \ref{fig:Example of singnals} as examples, the covariance matrices are calculated for varied time delays $\tau_k$, as shown in Fig. \ref{fig:toy problem}. By stacking the calculated covariance matrices, three-rank tensors  ${{{\bm{\mathcal{R}}}_{\mathbf{x}}}}$ (Eq. (\ref{eq:definition of tensor})) are constructed. The size of the tensor is equal to the numbers of sensors (10 $\sim$ 5 in this example) $\times$ numbers of sensors (10 $\sim$ 5 in this example) $\times$ numbers of time delays.

\begin{table}[H]
	\begin{center}
		\caption{Ground truth of natural frequencies and damping ratios by theoretical modal analysis.}
		\begin{tabular}{lcccccccccc}
			\hline
			Mode number&1&2&3&4&5&6&7&8&9&10\\
			\hline \hline
			Natural frequency [Hz] \\ (uniform mass and spring constants)&1.00&2.98&4.89&6.69&8.34&9.81&	11.1&12.1&12.8&13.2\\
			(non-uniform mass and spring constants)&2.41&3.86&5.43&7.04&8.58&9.96&	11.2&12.1&12.8&13.2\\
			\hline
			Damping ratio [\%] &5.00&1.68&1.02&0.750&0.600&0.510&0.450&0.410&0.390&0.380\\
			\hline
		\end{tabular}
	\end{center}
	\label{table:Ground truth_u}
\end{table}

\subsection{Identification of CP rank}
\begin{figure}[H]
	\begin{center}
		\begin{tabular}{c}
					\includegraphics[clip, width=0.9\linewidth]{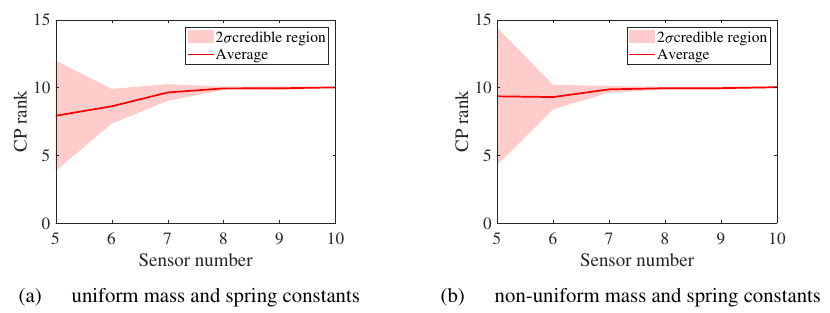}
		\end{tabular}
	\end{center}		
	\caption{Estimated CP rank (i.e. mode number). The CP rank is accurately determined through Bayesian CP decomposition using more than 7 sensors.}
	\label{fig:estimated CPrank}
\end{figure}

Average and confidence interval $2\sigma$ of 1000-times identification of CP ranks using each number of sensors are shown in Fig. \ref{fig:estimated CPrank}.
As the CP ranks correspond to the true numbers of vibration modes for 10, 9, and 8 sensors, the proposed OMA can function for under-determined systems. In contrast, although 7, 6, and 5 sensors theoretically satisfy the uniqueness of CP decomposition, the CP rank does not correspond to the true numbers of vibration modes. Therefore, more sensors are practically needed than the minimum numbers theoretically required to identify all vibration modes.

\subsection{Accuracy of identified modal parameters}
\begin{figure}[H]
\begin{center}
		\begin{tabular}{c}
\includegraphics[clip, width=1\linewidth]{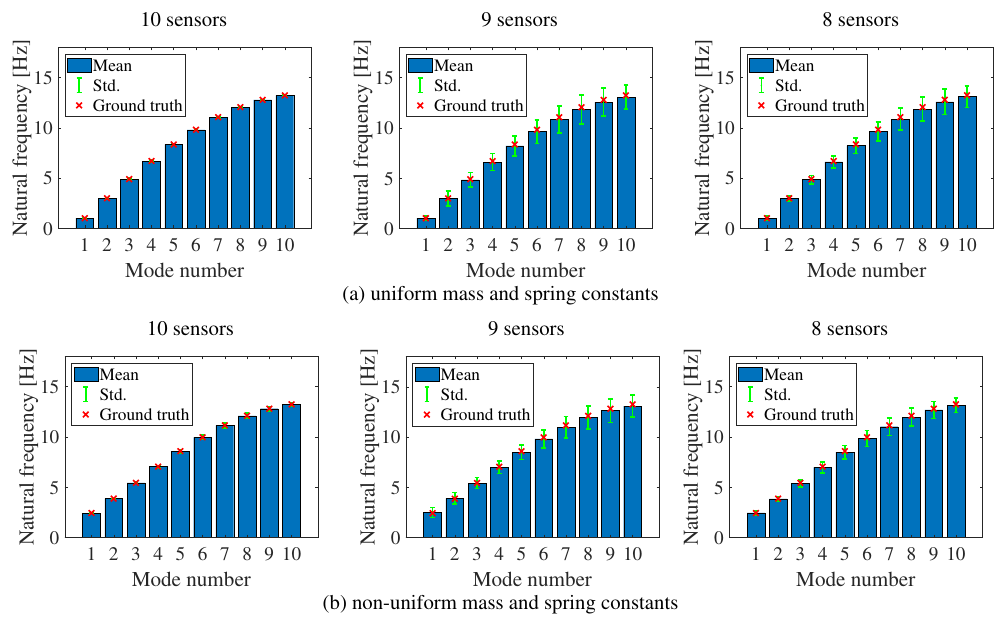}
	\end{tabular}
\end{center}		

\caption{Comparison between the estimated natural frequencies and ground truth. The natural frequencies are accurately estimated from under-determined systems.}
\label{fig:Fn_nu}
\end{figure}

\begin{figure}[H]
	\begin{center}
		\begin{tabular}{c}
					\includegraphics[clip, width=1\linewidth]{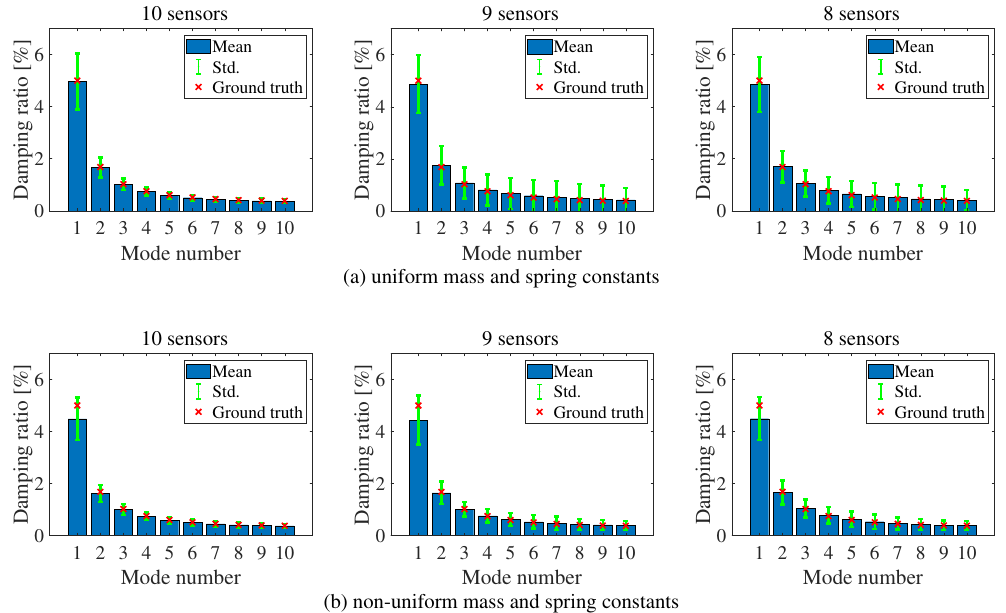}
		\end{tabular}
	\end{center}		

	\caption{Comparison between the estimated damping ratios and ground truth. Estimation of damping ratios is inaccurate regardless of sensor number.}
	\label{fig:Damping_nu}
\end{figure}

\begin{figure}[H]
	\begin{center}
		\begin{tabular}{c}
					\includegraphics[clip, width=0.8\linewidth]{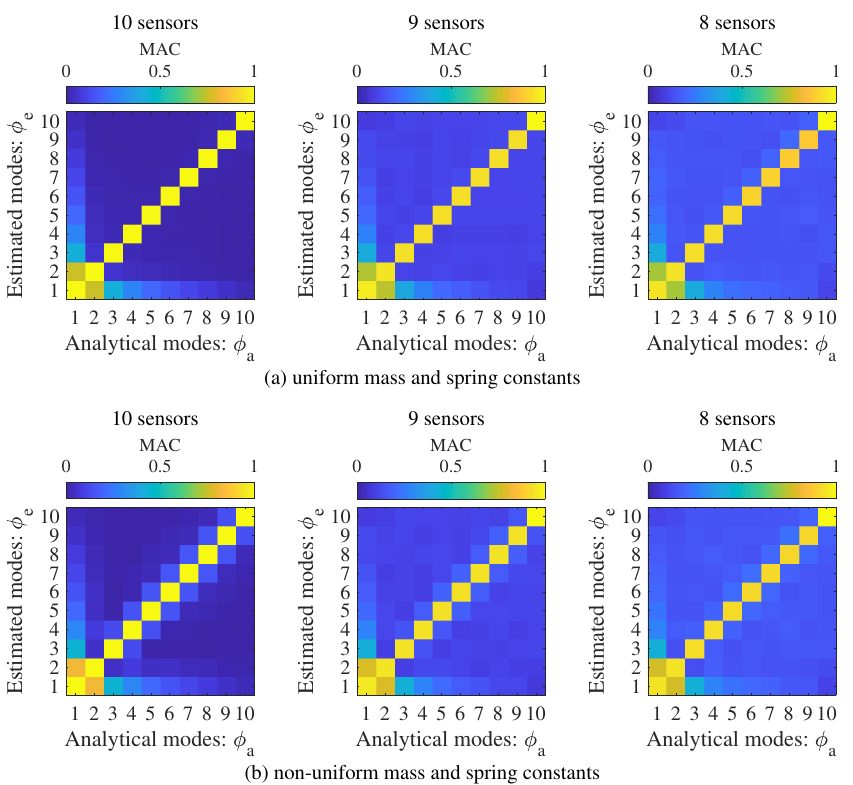}
		\end{tabular}
	\end{center}
	\caption{Average MAC between the estimated and analytical mode vectors. As diagonal components of MAC are dominant, the modes are appropriately separated using the Bayesian CP factorization.}
	\label{fig:MAC_nu}
\end{figure}

For the 10, 9, and 8 sensors that provide the true numbers of vibration modes in stable, the modal parameters are validated by comparing with the analytical solutions. First, the average and standard deviation of natural frequencies are shown in Fig. \ref{fig:Fn_nu}. Although the standard deviation of identified natural frequencies becomes larger as the number of sensors decreases, the error between the analytical and identified values is small. In contrast, the identification of damping ratios is not stable as shown in Fig. \ref{fig:Damping_nu} because the standard deviation of the identified damping ratios is large. Finally, the average of MAC between the analytical and identified modes is shown in Fig. \ref{fig:MAC_nu}, where the proposed OMA appropriately decomposes the vibration modes because the diagonal elements are dominant.

These numerical examples indicate that the proposed OMA can identify the natural frequencies and modal vectors from limited sensors without tuning the numbers of vibration modes before identification.
\vspace{0.5cm}
\newpage

\section{Conclusion}
This study proposes OMA via Bayesian CP decomposition of the three-rank tensors constructed by stacking the covariance matrices with varied time delays. Automatic determination of the numbers of vibration modes and appropriate identification of modal parameters realized as follows:

\vspace{0.5cm}
\begin{itemize}
	\item
	Considering that the uniqueness of CP decomposition is satisfied, the OMA can identify the modal parameters when the number of sensors is less than the number of vibration modes. Furthermore, the CP rank (number of natural frequencies) can be automatically determined via Bayesian CP decomposition. Therefore, we can identify the modal parameters from under-determined systems without tuning the numbers of vibration modes before identification.
	\vspace{0.5cm}
	\item
	As a result of numerical experiment by using 10-DOF mass-spring systems, the numbers of vibration modes can be automatically determined for under-determined systems (the numbers of sensors are less than numbers of vibration modes.). In addition, the identified natural frequencies and modal vectors correspond to the analytical solutions. We demonstrated that the proposed OMA functions appropriately via numerical experiments.
\end{itemize}

\end{document}